\documentclass[twoside, epsfig, a4paper]{article}

\input oejv.sty

\usepackage{amsmath}
\usepackage{amssymb}
\usepackage[dvips]{graphicx}
\usepackage{subfig}

\begin{document}

\OEJVhead{March 2009} \OEJVtitle{Three New Variable Stars in
Indus} \OEJVauth{Alex Golovin$^1$, Kirill Sokolovsky$^{2, 3}$,
Natalia Virnina$^{4, 5}$, Javier L{\'o}pez Santiago$^6$}
\OEJVinst{Main Astronomical Observatory of National Academy of
Sciences of Ukraine, Kiev, UKRAINE \\ {\tt
golovin.alex@gmail.com}} \OEJVinst{Max-Planck-Institute f\"ur
Radioastronomie, Bonn, GERMANY \\ {\tt
ksokolov@mpifr-bonn.mpg.de}} \OEJVinst{Astro Space Center of
Lebedev Physical Institute, Moscow,
RUSSIA} \OEJVinst{Odessa National University, Odessa, UKRAINE \\
{\tt virnina@gmail.com} \\\OEJVinst{Odessa National Maritime
University, Odessa, UKRAINE} \OEJVinst{Physic Faculty,
Universidad Complutense de Madrid,  Madrid, SPAIN\\
{\tt jls@astrax.fis.ucm.es}}}

\OEJVabstract{We report the discovery of three new variable stars
in Indus: USNO-B1.0 0311-0760061, USNO-B1.0 0309-0771315, and
USNO-B1.0 0315-0775167. Light curves of 3712 stars in a $87'
\times 58'$ field centered on the asynchronous polar CD Ind were
obtained using a remotely controlled 150 mm telescope of Tzec Maun
Observatory (Pingelly, Western Australia). The VaST software based
on SExtractor package was used for semi-automatic search for
variable stars. We suggest the following classification for the
newly discovered variable stars: USNO-B1.0 0311-0760061 - RR
Lyr-type, USNO-B1.0 0309-0771315 - W UMa-type, and USNO-B1.0
0315-0775167 - W UMa-type.}

\begintext
\section{Introduction}
We observed a  field centered on the asynchronous polar CD Ind
(magnetic cataclysmic binary; see \citeauthor{CDInd1},
\citeyear{CDInd1} and \citeauthor{CDInd2}, \citeyear{CDInd2}) from
December 21, 2008 until January 19, 2009 during 10 nights.
Observations were done remotely at Tzec Maun Observatory
(Pingelly, Western Australia) using the Takahashi TOA-150
apochromatic refractor (D=150 mm, F = 1095 mm) and SBIG STL-6303
CCD camera. All observations were conducted with the Bessel R
filter. The field of view was $87' \times 58'$. About 4 000
objects were detected on each frame. Our goal was to find new
variable stars.

\section{Data Analysis}
\begin{figure}[h!]
\begin{center}
\includegraphics [width=14cm]{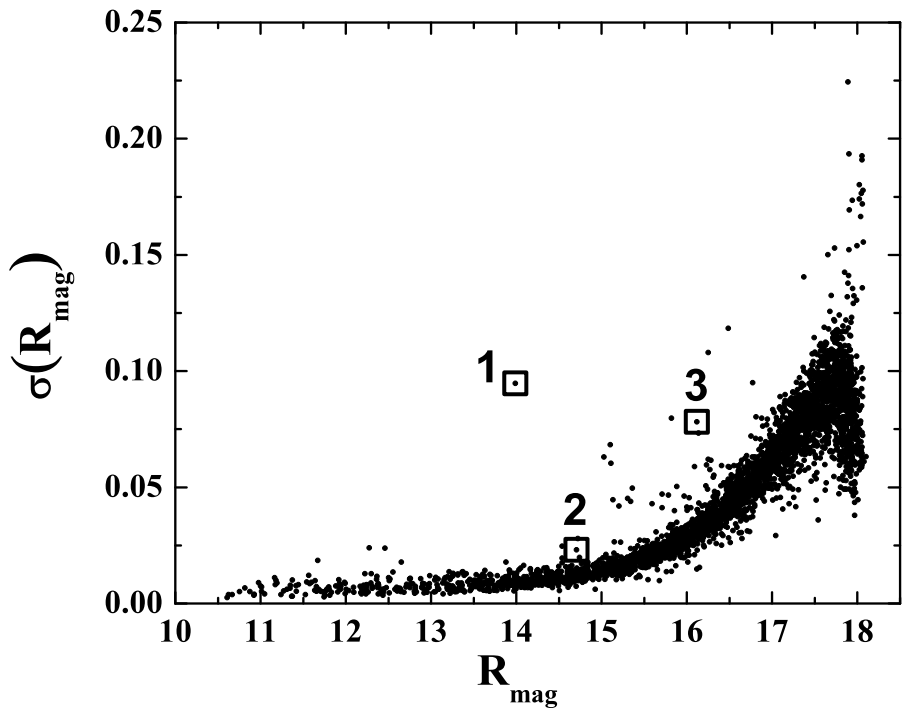}
\caption{\emph{RMS}-scatter vs. mean magnitude diagram.}
\label{sigma}
\end{center}
\end{figure}

To carry out the search for new variable stars we analyzed 108 CCD
frames. The VaST software (developed by K. Sokolovsky and A.
Lebedev, described by \citeauthor{VaST}, \citeyear{VaST}) based on
the SExtractor routine \citep{s1} was used to obtain light curves
of 3712 objects in the vicinity of CD Ind. The $R_{mag}$ range was
$10.\!\!^{\rm m}6-18.\!\!^{\rm m}1$. To identify possible variable
stars among them, we created a \emph{RMS}-scatter versus mean
magnitude diagram (see Fig. \ref{sigma}). The fainter objects tend
to have larger scatter. Those objects, which bounce from this
relation, (objects with \emph{RMS}-scatter significantly larger
than the typical value for their mean magnitude)           are
good candidates to be variable stars. False positives could also
be caused if object is a galaxy or blended, either by the presence
of close companion or image defect etc.

Three previously unknown variable stars were discovered on our CCD
images. They are marked by squares on Fig. \ref{sigma}, where 1
corresponds to USNO-B1.0 0311-0760061, 2 - to USNO-B1.0
0309-0771315, and 3 - to USNO-B1.0 0315-0775167.

The instrumental magnitudes were scaled to $R_{mag}$ with the
reference star USNO-B1.0 0313-0766011 ($\alpha_{2000} = 21^{\rm h}
15^m 36.\!\!^{\rm s}85, \delta_{2000} = -58^{\circ}
41'52.\!\!^{\prime\prime}7 $). In USNO-B1.0 catalog it has
magnitudes $R_1 = 12.\!\!^{\rm m}42$ and $R_2 = 12.\!\!^{\rm
m}43$. All coordinates in this paper were taken from the USNO-B1.0
catalog \citep{Monet}.

The light curves of these newly discovered variable stars were
searched for periodic variation of brightness using the
``Period04'' package, developed by Patrick Lenz
(\citeauthor{period04}, \citeyear{period04}). A discrete Fourier
transform (DFT) algorithm was used for the statistical analysis.
Notably, dates are heliocentrically corrected. The average zero
point (object mean magnitude) was subtracted to prevent the
appearance of spurious features on a periodogram centered at
frequency 0.0. We calculated a periodogram with frequency range
from 0 to 50 cycles per day ($c \cdot d^{-1}$) and steps of
0.00172 $c \cdot d^{-1}$.

We also looked for X-ray counterparts of these three stars.
Unfortunately, no \emph{XMM-Newton} or \emph{Chandra} observations
(either publicly released or not) have included this field.
Similarly, no \emph{ROSAT}/HRI observation was pointed in the
direction of our targets. The stars were not detected in the
\emph{ROSAT} All Sky Survey, which is complete down to 0.03 cnt/s.
Thus, we assume a maximum X-ray flux in the \emph{ROSAT}/PSPC
energy band (0.1-2.4 keV) of $\sim 7 \times 10^{-13}$ erg s$^{-1}$
cm$^{-2}$ (for radiation spectrum of optically thin plasma with a
mean temperature $kT = 0.5$ keV = 5.8$\cdot 10^6$ K).

\section{Notes on Individual Objects}

\subsection{USNO-B1.0 0311-0760061}

\begin{figure}[h!]
\begin{center}
\includegraphics [width=7cm]{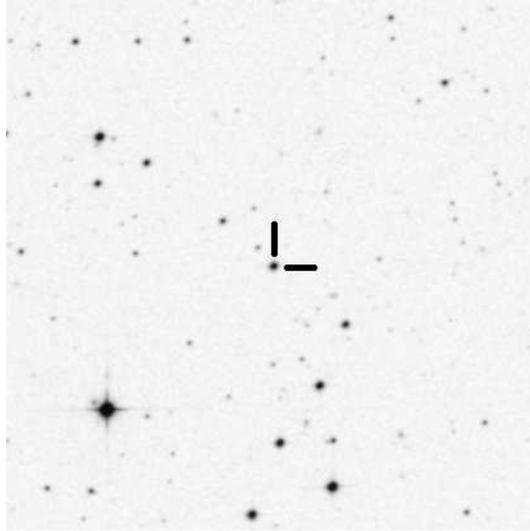}
\caption{$7'\times7'$ DSS image of the vicinity of USNO-B1.0
0311-0760061.} \label{var4456map}
\end{center}
\end{figure}

USNO-B1.0 0311-0760061 (= GSC 08805-00255, $\alpha_{2000} = 21^{h}
11^{m} 35.\!\!^{\rm s}32$, $\delta_{2000} = -58^{\circ}
50^\prime14.\!\!^{\prime\prime}3$) displays $0.\!\!^{\rm m}27$
variability with a mean level of brightness of $13.\!\!^{\rm
m}99$. Fig. \ref{var4456map} shows $7'\times7'$ DSS image
(POSS2/UKSTU Red plate, hereafter) of the vicinity of this star.
North is up, east is left.

The object is listed in the UCAC2 Catalogue \citep{ucac2, ucac1}
with proper motion values of $pmRA = -2.9$ mas/yr and $pmDE =
-7.6$ mas/yr (mas/yr - milli-second of arc per year). PPMX
Catalogue \citep{ppmx2, ppmx1} gives values of $pmRA = 1.43$
mas/yr and $pmDE = -5.55$ mas/yr.

\begin{figure}[h!]
\centering \subfloat[]{\label{4456fou}\includegraphics
[width=0.495\textwidth]{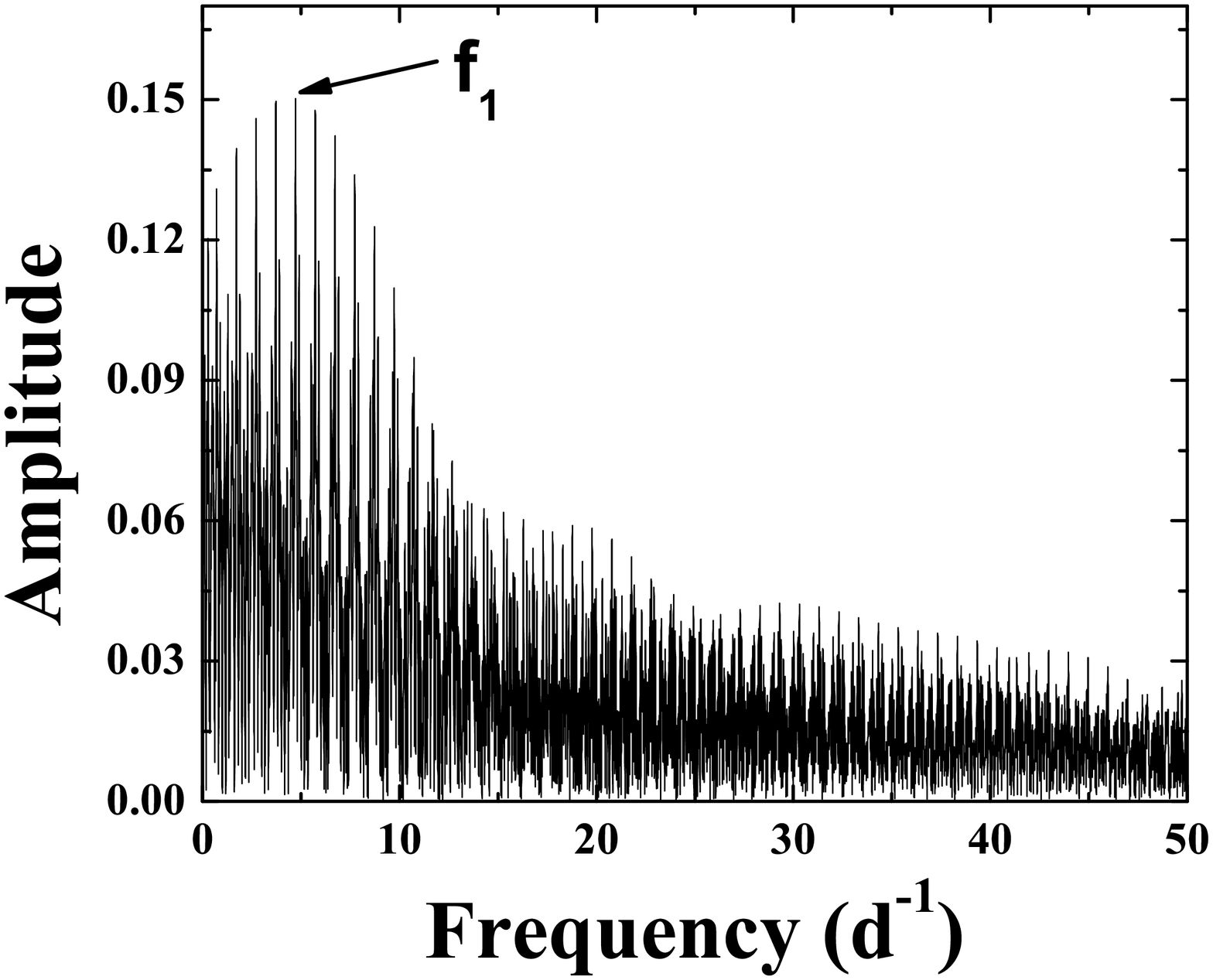}}
\subfloat[]{\label{4456lc}\includegraphics
[width=0.495\textwidth]{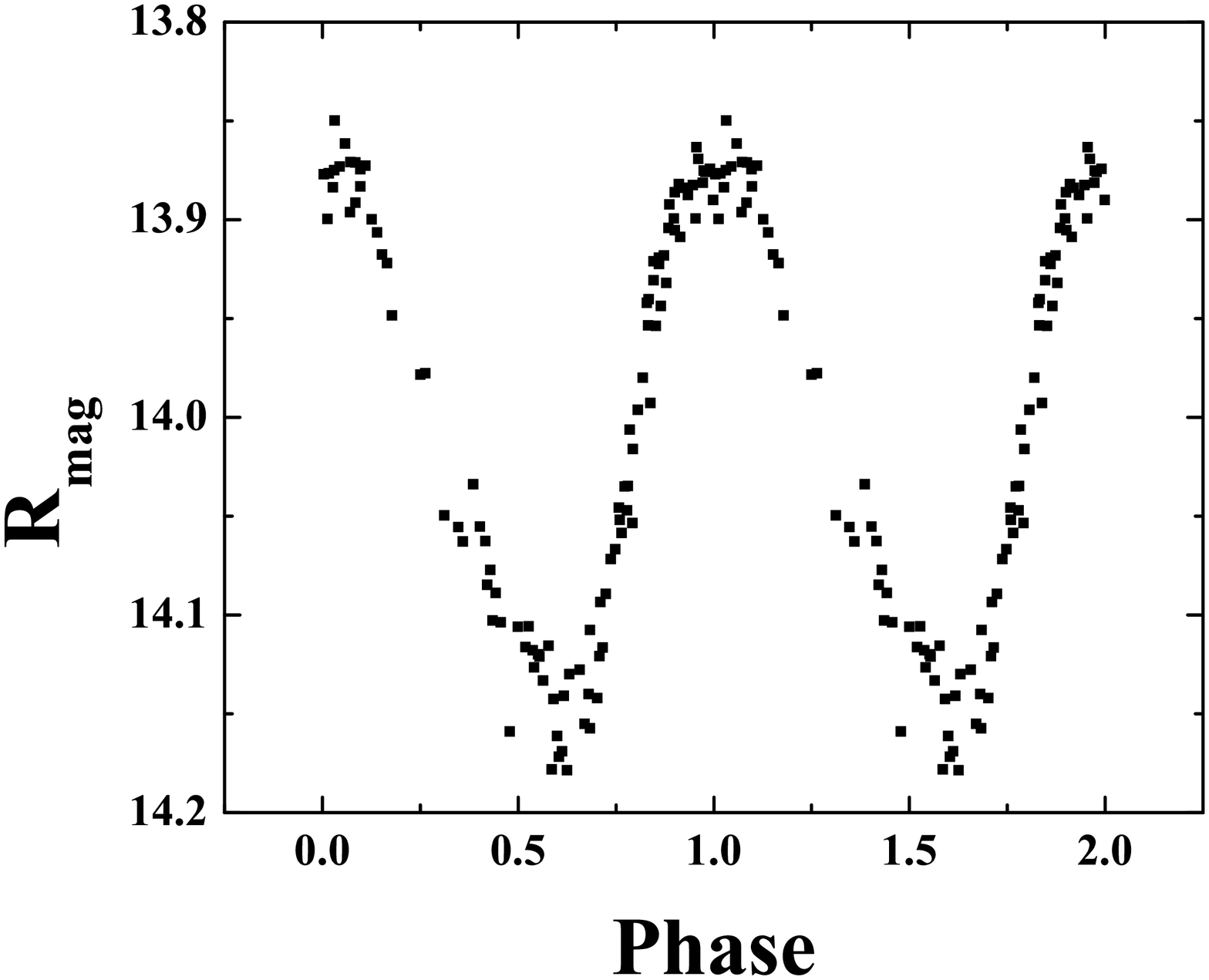}} \caption{Periodogram (a) and
phase plot (b) for USNO-B1.0 0311-0760061}\end{figure}

The dominant frequency in the DFT-periodogram $f_1 = 4.7213 \pm
0.0005$ $c \cdot d^{-1}$ implies a period of $P = 0.\!\!^{\rm
d}2118$ (Fig. \ref{4456fou}).

The following ephemeris were used for the phase plot
(Fig.~\ref{4456lc}):
$$HJD_{Max}  =  2454823.1005 + 0.\!\!^{\rm d}2118 \cdot E$$
The number of significant digits in the ephemeris reflects an
estimated precision. Time in ephemeris hereinafter is UTC.

Based on the light curve, we classify this object as an RRC-type
pulsating star. Rising time (M-m; time elapsed from minimum to the
nearest further maximum) estimated to be $48\%$ (expressed in
percentage of the period).

The pulsation period of the star lies just outside the boundary of
delta Scuti-type variables. According to \citet{Poretti} virtually
all pulsating variables with $0.\!\!^{\rm d}20 < P < 0.\!\!^{\rm
d}25$ are RRC-type. The smooth, sinusoidal shape of the light
curve also helps to avoid confusion with High Amplitude
$\delta$-Scuti stars (HADS) which usually have asymmetric light
curves with a steep ascending branch and a sharp maximum.

\subsection{USNO-B1.0 0309-0771315}
\begin{figure}[h!]
\begin{center}
\includegraphics [width=7cm]{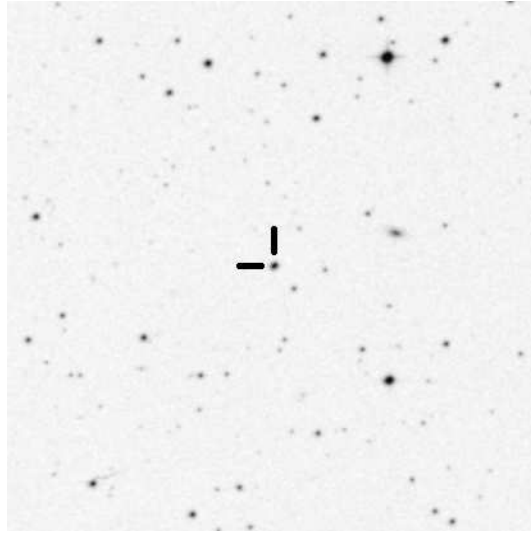}
\caption{$7'\times7'$ DSS image of the vicinity of USNO-B1.0
0309-0771315.} \label{var2998map}
\end{center}
\end{figure}

USNO-B1.0 0309-0771315 ($\alpha_{2000}=21^{\rm h} 16^m
20.\!\!^{\rm s}28, \delta_{2000} = -59^{\circ}
01'56.\!\!^{\prime\prime}7$) is the next star with previously
unknown variability. Fig. \ref{var2998map} shows the DSS image of
its vicinity with the variable being marked. The proper motion of
this object is $pmRA = -7.9$ mas/yr, $pmDE = 8.2$ mas/yr
\citep{ucac2}.

\begin{figure}[h!]
\centering \subfloat[]{\label{2998fou}\includegraphics
[width=0.5\textwidth]{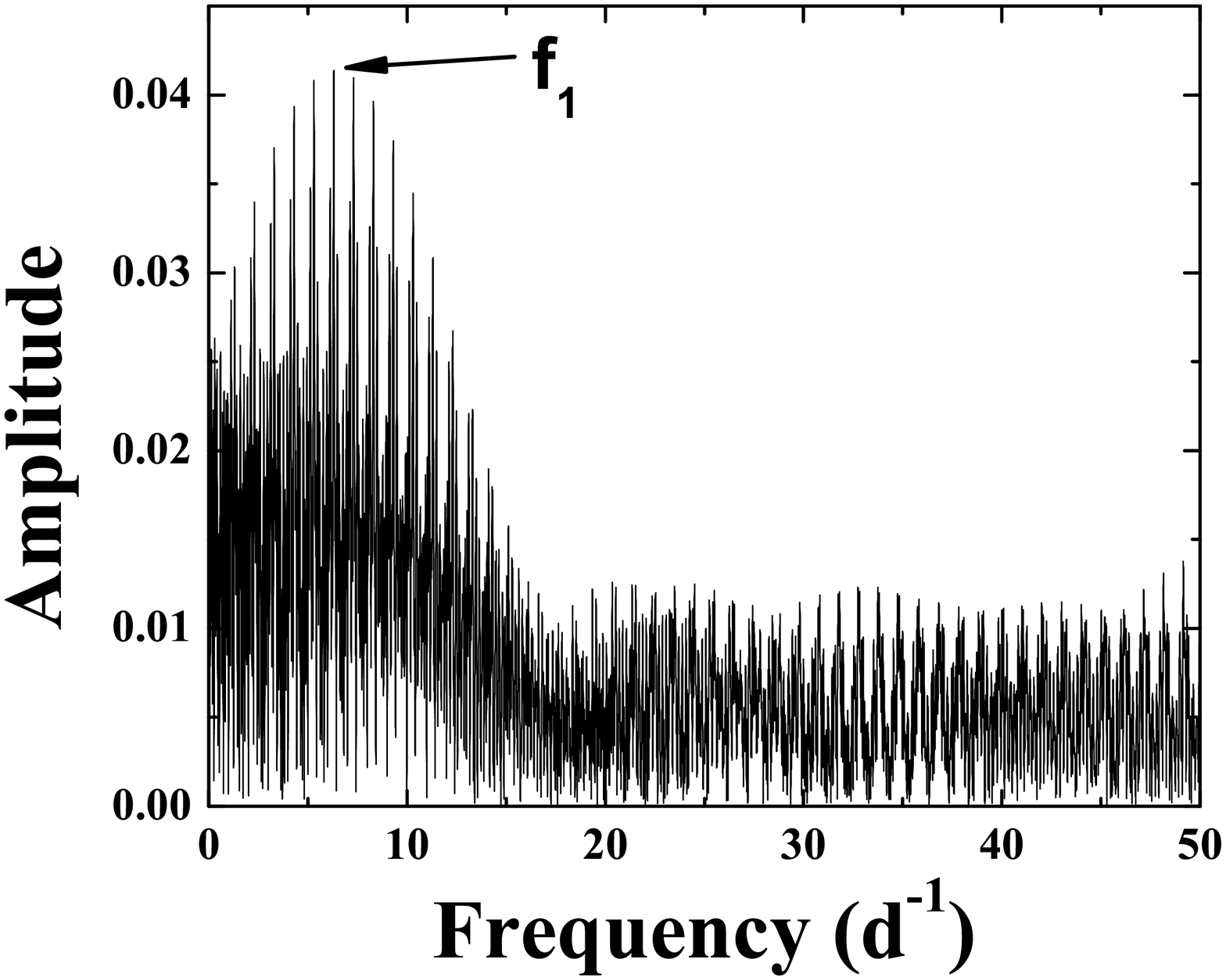}}
\subfloat[]{\label{2998lc1}\includegraphics
[width=0.495\textwidth]{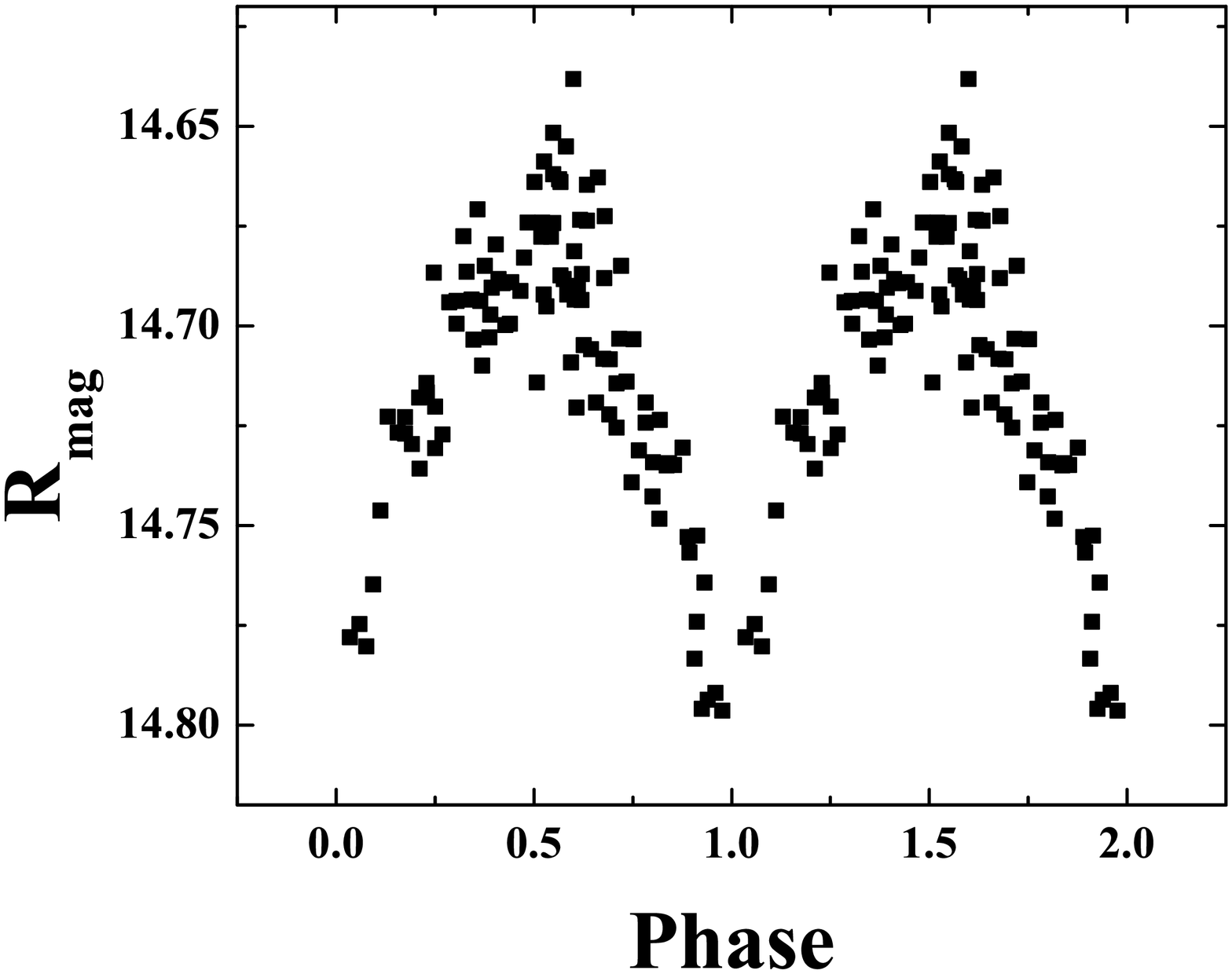}}
\subfloat[]{\label{2998lc2}\includegraphics
[width=0.495\textwidth]{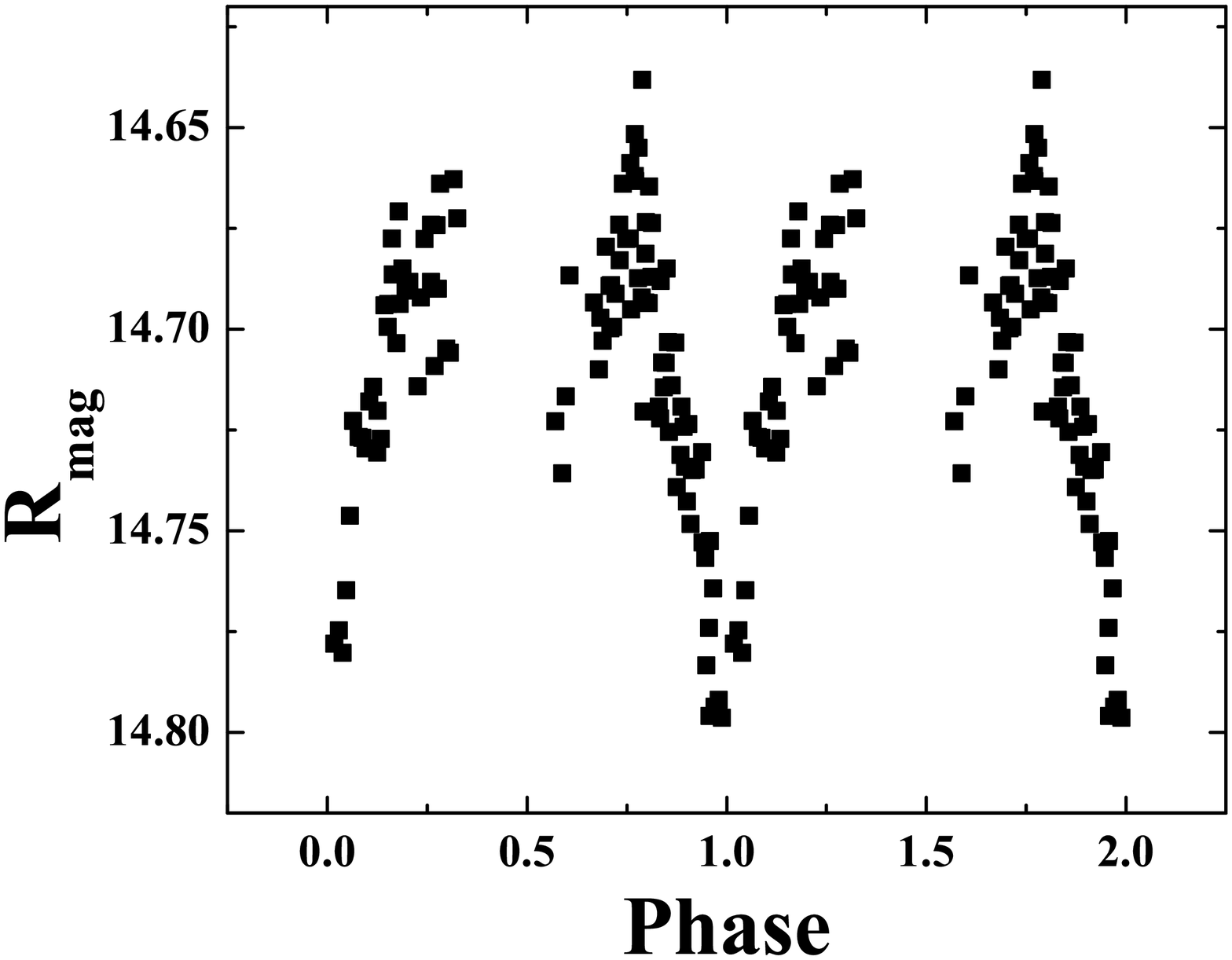}} \caption{Periodogram (a) and
phase plots for USNO-B1.0 0309-0771315 folded with P-value of the
period (b) and $2 \cdot P_1$ (c)}\end{figure}

The mean brightness of the star is $14.\!\!^{\rm m}73$, with an
amplitude variations of $0.\!\!^{\rm m}12$. Our periodogram
analysis (Fig. \ref{2998fou}) revealed the dominant frequency to
be $f_1 = 6.3102 \pm  0.0014$ $c \cdot d^{-1}$ ($P_1 = 0.\!\!^{\rm
d}1585$), although a twice longer period could not be excluded
because of insufficient photometric coverage, so we folded our
data with $P_1$ and $2 \cdot P_1$ (Fig. \ref{2998lc1},
\ref{2998lc2}). Most probably, the represented light curve  could
be explained if this star is a W UMa-type eclipsing binary.
Ephemeris (with $2 \cdot P_1$) are:
$$HJD_{Min}  =  2454823.0551 + 0.\!\!^{\rm d}3169 \cdot E$$

\subsection{USNO-B1.0 0315-0775167}

USNO-B1.0 0315-0775167 ($\alpha_{2000} = 21^{\rm h} 20^m
26.\!\!^{\rm s}46, \delta_{2000} =  -58^{\circ}
25'01.\!\!^{\prime\prime}7$) shows $0.\!\!^{\rm m}25$ variability,
with a mean level of brightness of $16.\!\!^{\rm m}10$ in R-band.

\begin{figure}[h!]
\begin{center}
\includegraphics [width=7cm]{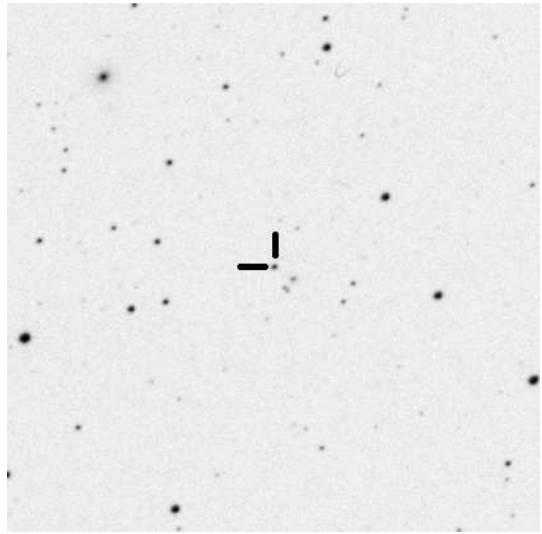}
\caption{$7'\times7'$ DSS image of the vicinity of USNO-B1.0
0315-0775167.} \label{var1098map}
\end{center}
\end{figure}

Fig. \ref{var1098map} shows $7'\times7'$ DSS image of the
USNO-B1.0 0315-0775167 vicinity with the variable star being
marked.

\begin{figure}[h!]
\centering \subfloat[]{\label{1098fou}\includegraphics
[width=0.5\textwidth]{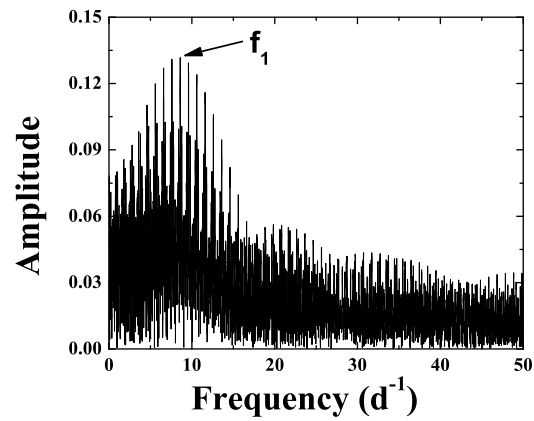}}
\subfloat[]{\label{1098lc1}\includegraphics
[width=0.495\textwidth]{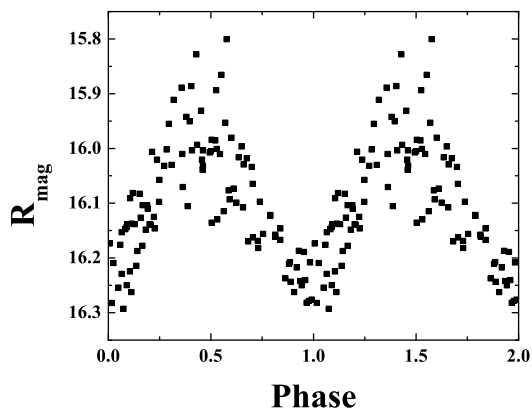}}
\subfloat[]{\label{1098lc2}\includegraphics
[width=0.495\textwidth]{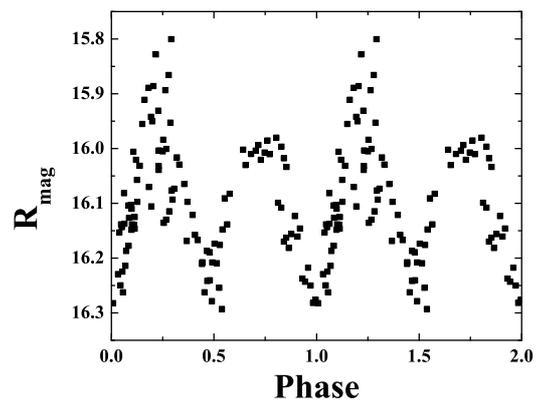}} \caption{Periodogram (a) and
phase plots for USNO-B1.0 0315-0775167 folded with P-value of the
period (b) and $2 \cdot P_1$ (c)}\end{figure}

Periodogram analysis (Fig. \ref{1098fou}) revealed a dominant peak
at a frequency $f_1 = 8.5987 \pm 0.0013$ $c \cdot d^{-1}$ $(P_1 =
0.\!\!^{\rm d}1163)$. As in the case of USNO-B1.0 0309-0771315, we
assumed $2 \cdot P_1$ to be the real period of the system (Fig.
\ref{1098lc1}, \ref{1098lc2}). From the inspection of the light
curve we suspect this star is a W UMa-type eclipsing binary,
although further observations are required. Ephemeris (with $2
\cdot P_1$) are:
$$HJD_{Min}  =  2454823.1042 + 0.\!\!^{\rm d}2326 \cdot E$$

The light curve of USNO-B1.0 0315-0775167 (Fig. \ref{1098lc2}) has
a large scatter at the maximum which follows primary minimum
compared to the maximum which follows secondary minimum. This is,
most probably, an instrumental effect. However, since the exact
nature of this effect is not clear, intrinsic stellar variability
can not be completely discarded.

It is sometimes difficult to distinguish RRC-type variables from
EW-type eclipsing binaries using only light curves since both
types have nearly symmetrical light curves and comparable periods.
However after close visual inspection of light curves, it is often
possible to distinguish between them using the following features:
\begin{itemize}
    \item  minima of EW-type variables are usually slightly sharper than
maxima;
    \item  EW-type binaries may have minima of slightly different
    depth;
    \item  EW-type binaries may have slightly different brightness at
maximum light before and after the primary minimum (O'Connel
effect);
    \item  RRC-type variables may have slightly asymmetric
light curves with the ascending branch a little steeper then
descending branch.
\end{itemize}

Bearing all these features in mind,  it is  possible to
 distinguish between RRC- and EW-type variables using an accurate light curve.

 Sometimes it is also possible to use
color information to distinguish between RR Lyrae variables which
are A-F type giants and W Ursae Majoris variables which usually
have spectral types later then F.

We analyzed 2MASS data (JHK magnitudes with errors and Julian date
of observation) of the studied objects. Last column in table
\ref{table:2MASS} shows spectral type, estimated from JHK colour
indexes according to \citet{2mass}.

The Two Micron All-Sky Survey (2MASS) is a single-epoch survey so
we have only one data point in each color. However, the good thing
about the 2MASS is that JHK colors are measured truly
simultaneously using a beam splitter, due to that fact the 2MASS
colors are not affected by the variability. We neglect the small
color change which may be a function of the variability phase
because we are interested only in rough spectral classification,
and, anyway, only rough spectral classification is possible from
the broad-band photometry.

\begin{table}[h!]
\begin{center}
\caption{\label{table:2MASS} 2MASS data for studied objects}
\begin{tabular}{lllllllll}
\hline
Object & $J_{mag}$ & $e_{J_{mag}}$ & $H_{mag}$ & $e_{H_{mag}}$ & $K_{mag}$ & $e_{K_{mag}}$ & JD & Sp \\
\hline
2MASS   21113531-5850146 & 13.702 & 0.026 & 13.632 & 0.041 & 13.485 & 0.048 & 2451490.5454 & $\sim A5$ \\
2MASS   21162028-5901569 & 14.307 & 0.041 & 13.980 & 0.058 & 13.983 & 0.060 & 2451490.5730 & $\sim G $\\
2MASS   21202644-5825020 & 15.199 & 0.052 & 14.826 & 0.077 & 15.115 & 0.133 & 2451490.5823 & $\sim G$ \\
\hline
\end{tabular}
\end{center}
\end{table}

\begin{table}[h!]
\begin{center}
\caption{\label{table:summary} Summarized Data on discovered
variable stars}
\begin{tabular}{llllllll}
\hline
Object & $\alpha_{2000}$ & $\delta_{2000}$ & $R_{Max}$ & $R_{min}$ & P & $E_0$ (UTC) & Type \\
\hline USNO-B1.0 0311-0760061 & $21^{h} 11^{m} 35.\!\!^{\rm s}32$
& $-58^{\circ}
50^\prime14.\!\!^{\prime\prime}3$ & $13.\!\!^{\rm m}86$ & $14.\!\!^{\rm m}13$ & $0.\!\!^{\rm d}2118$ & 2454823.1005 & RRC \\
USNO-B1.0 0309-0771315 & $21^{\rm h} 16^m 20.\!\!^{\rm s}28$ &
$-59^{\circ}
01'56.\!\!^{\prime\prime}7$ & $14.\!\!^{\rm m}67$ & $14.\!\!^{\rm m}79$ & $0.\!\!^{\rm d}3169$ & 2454823.0551 & EW:  \\
USNO-B1.0 0315-0775167 & $21^{\rm h} 20^m 26.\!\!^{\rm s}46$ &
$-58^{\circ} 25'01.\!\!^{\prime\prime}7$ & $15.\!\!^{\rm m}98$ &
$16.\!\!^{\rm
m}23$ & $0.\!\!^{\rm d}2326$ & 2454823.1042 & EW: \\
\hline
\end{tabular}
\end{center}
\end{table}

\newpage


\newpage
{\color{white}.}

\section{Conclusions}

We examined light curves of 3712 objects in the $87' \times 58'$
field centered on CD Ind and discovered three new variable stars.
Two of them are most probably  W UMa-type eclipsing binaries (EW),
and the other one is most probably an  RR Lyr-type (RRC) star. We
show that \emph{RMS-scatter vs. Magnitude} diagram is a useful
tool to discover new variable stars.

Table \ref{table:summary} summarizes parameters of discovered
variable stars, such as USNO-B1.0 designation, equatorial
coordinates, brightness in maximum and minimum, period (in
fractions of days), zero-epoch ($E_0$, time of maximum for RRC
star and time of minimum for EW stars) and GCVS type of
variability.

\section*{Acknowledgments} This research is based on data collected with the Tzec Maun
Observatory, operated by the Tzec Maun Foundation. Special thanks
to Ron Wodaski (director of the observatory) and Donna
Brown-Wodaski (director of the Tzec Maun Foundation). The authors
are thankful to Leonid Elenin (``Astrogalaxy'' team leader) for
making these observations possible. JLS acknowledges financial
support by the ASTROCAM project S-0505/ESP/000237. This work is
partially supported by the Russian-Ukrainian grant F28.2/081 of
FFI.

We would like to thank Alexander Pushkarev, Elena Pavlenko,
Lyudmila Pakuliak, Tatyana Sergeeva, Mislav Balokovic and Yuliana
Kuznyetsova for reviewing this manuscript. Special thanks to Aaron
Price and David Trethewey for their help in preparing this paper.

This publication has made use of the Aladin interactive sky atlas,
operated at CDS, Strasbourg, France and the International Variable
Star Index (VSX) operated by the AAVSO. K. Sokolovsky was
supported by the International Max Planck Research School (IMPRS)
for Astronomy and Astrophysics at the universities of Bonn and
Cologne. This research has made use of NASA's Astrophysics Data
System. This research has made use of data products from the Two
Micron All Sky Survey, which is a joint project of the University
of Massachusetts and the Infrared Processing and Analysis
Center/California Institute of Technology, funded by the National
Aeronautics and Space Administration and the National Science
Foundation.

\end{document}